\documentclass[11pt]{article}
\usepackage[utf8]{inputenc}
\usepackage{amsfonts}
\usepackage{amsmath}
\usepackage{amssymb}
\usepackage{indentfirst}
\usepackage{graphicx}
\usepackage[colorlinks]{hyperref}
\usepackage{cite}
\usepackage{colortbl}
\usepackage{bm}
\usepackage{microtype}
\usepackage{caption}
\usepackage[normalem]{ulem}

\numberwithin{equation}{section}

\allowdisplaybreaks
\begin{document}

\begin{titlepage}
\vskip 4cm

\begin{center}
\textbf{\LARGE 3D Carrollian gravity from 2D Euclidean symmetry}
\par\end{center}{\LARGE \par}

\begin{center}
	\vspace{1cm}
	\textbf{Patrick Concha}$^{\ast, \bullet}$,
	\textbf{Evelyn Rodríguez}$^{\ast, \bullet}$,
   \textbf{Sebastián Salgado}$^{\star}$
	\small
	\\[6mm]
	$^{\ast}$\textit{Departamento de Matemática y Física Aplicadas, }\\
	\textit{ Universidad Católica de la Santísima Concepción, }\\
\textit{ Alonso de Ribera 2850, Concepción, Chile.}
	\\[3mm]
 $^{\bullet}$\textit{Grupo de Investigación en Física Teórica, GIFT, }\\
	\textit{ Universidad Católica de la Santísima Concepción, }\\
\textit{ Alonso de Ribera 2850, Concepción, Chile.}
\\[3mm]
	$^{\star}$\textit{Instituto de Alta Investigación, Universidad de Tarapac\'{a},\\ Casilla 7D, Arica, Chile} \\[5mm]
	\footnotesize
	\texttt{patrick.concha@ucsc.cl},
	\texttt{erodriguez@ucsc.cl},
 	\texttt{ssalgador@gestion.uta.cl}
 
	\par\end{center}
\vskip 20pt
%\begin{abstract}
\centerline{{\bf Abstract}}
\medskip
\noindent Carroll symmetry arises from Poincaré symmetry when the speed of light is sent to zero. In this work, we apply the Lie algebra expansion method to find the Carroll versions of different gravity models in three space-time dimensions. Our starting point is the 2D Euclidean AdS algebra along with its flat version. Novel and already known Carrollian algebras, such as the AdS-Carroll and Carroll-Galilei ones are found, and the Chern--Simons gravity theories based on them are constructed. Remarkably, after the expansion, the vanishing cosmological constant limit applied to the 2D Euclidean AdS algebra converts into a non-relativistic limit in three space-time dimensions. We extend our results to Post-Carroll-Newtonian algebras which can be found by expanding a family of 2D Euclidean algebras. 

%\end{abstract}
\end{titlepage}\newpage {\baselineskip=12pt \tableofcontents{}}

\section{Introduction}

Symmetries are essential for analyzing and understanding several aspects of physical theories. In this context, ultra-relativistic (UR) symmetries, also referred to as Carroll symmetries, have gained increasing attention in recent years. These symmetries emerge in the ultra-relativistic limit as the speed of light, $c$, approaches zero \cite{LevyLeblond,Bacry:1968zf}. The Carroll group, introduced by Lévy-Leblond, is derived from the UR contraction of the Poincaré group which is the dual of the non-relativistic (NR) contraction. Such contraction is dual to the non-relativistic (NR) one (i.e., when $c$ tends to infinity), leading to the Galilean group. Originally, the Carroll symmetry was presented as a peculiar limit of the Poincaré one\cite{LevyLeblond} due to the absence of causality in a Carroll Universe. Nowadays, models exhibiting Carroll symmetries have been explored in the context of tachyon condensation \cite{Gibbons:2002tv}, warped conformal field theories \cite{Hofman:2014loa}, and tensionless (super)strings \cite{Bagchi:2013bga, Bagchi:2018wsn}, asymptotic symmetries \cite{Perez:2021abf,Perez:2022jpr,Fuentealba:2022gdx}, flat holography \cite{Duval:2014uva,Hartong:2015xda,Hartong:2015usd,Bagchi:2016bcd,Donnay:2022aba,Saha:2022gjw,Saha:2023hsl,Nguyen:2023vfz,Bagchi:2023cen,Donnay:2023mrd}, and black hole horizon \cite{Donnay:2019jiz,Ciambelli:2019lap,Grumiller:2019fmp,deBoer:2023fnj,Ecker:2023uwm}.

Carroll theories have also been explored in the context of gravity, where two distinct versions,  denoted as electric Carroll gravity \cite{Henneaux:1979vn,Pekar:2024ukc} and magnetic Carroll gravity \cite{Bergshoeff:2017btm,Henneaux:2021yzg,Campoleoni:2022ebj} have been presented. Matter coupling and conformal approach to Carroll gravity have also been discussed in \cite{Hansen:2021fxi,Baiguera:2022lsw,Bergshoeff:2024ilz}. In three space-time dimensions, Carroll gravity can be expressed as a CS action \cite{Matulich:2019cdo} which can be derived from the $\mathfrak{iso}\left(2,1\right)$ CS action after a proper contraction. Generalizations of the Carroll CS gravity together with its supersymmetric extension can be found in \cite{Ravera:2019ize,Ali:2019jjp,Gomis:2019nih,Concha:2021jnn,Concha:2024dap}. More recently, the Carroll CS gravity theories have been alternatively recovered from relativistic symmetries using the expansion method based on semigroups ($S$-expansion)\cite{Izaurieta:2006zz}\footnote{For further literature on expansion method applied to non-Lorentziam realm, see e.g.\cite{deAzcarraga:2019mdn,Bergshoeff:2019ctr,Concha:2019lhn,Penafiel:2019czp,Romano:2019ulw,Bergshoeff:2020fiz,Kasikci:2020qsj,Fontanella:2020eje,Concha:2022muu,Caroca:2022byi,Concha:2022you,Concha:2022jdc,Concha:2023bly}.}. The $S$-expansion is a useful tool not only to derive new Lie (super)algebras but also to obtain their corresponding invariant tensors, which are key ingredients when constructing CS actions. However, it must be noted that well-defined CS actions require the non-degeneracy of such tensors. Although the Carrollian regime does not suffer from degeneracy its Carrollian-Galilean version\footnote{Initially denoted as para-Galilei algebra in \cite{Bacry:1968zf}.} requires central charges to prevent it \cite{Concha:2023bly}.

In this work, motivated by the relevance of the Carrollian symmetries in diverse areas of physics, and the simplicity of 3D gravity models, we present an alternative approach to derive the 3D AdS-Carroll CS gravity and its Carroll-Galilean regime by expanding the 2D Euclidean AdS and Poincaré algebras. Intriguingly, we show that the vanishing cosmological constant limit $\Lambda\rightarrow 0$ applied to the 2D Euclidean AdS algebra can be interpreted as a non-relativistic limit $c\rightarrow\infty$ in the 3D sector. In particular, the flat limit allows us to obtain the Carroll-Galilei algebra from the AdS-Carroll one. However, both 2D Euclidean Poincaré and Carroll-Galilei algebras suffer from degeneracy. 

In the second part of our work, we deal with the degeneracy issue. As it was noticed in \cite{Matulich:2019cdo} and according to the Medina
and Revoy theorem \cite{Medina,Figueroa-OFarrill:1995opp}, one way to overcome such difficulty is to consider the most natural extension of the Poincaré algebra admitting a non-degenerate invariant metric in arbitrary space-time dimensions: The Maxwell algebra \cite{Bacry:1970ye,Bacry:1970du,Schrader:1972zd,Gomis:2017cmt}. In particular, we show that the expansion applied to the 2D Euclidean Maxwell algebra reproduces the 3D extended Carroll-Galilei algebra which admits a non-degenerate bilinear invariant trace. In higher space-time dimensions, the Maxwell symmetry and its generalizations have been widely explored in \cite{deAzcarraga:2010sw,Salgado:2014jka,Hoseinzadeh:2014bla,Concha:2014xfa,Concha:2014tca,Penafiel:2017wfr,Ravera:2018vra,Concha:2018zeb,Concha:2018jxx,Chernyavsky:2020fqs,Kibaroglu:2020tbr,Caroca:2021bjo,Matulich:2023xpw}. We end our work by introducing a new family of post-Carroll-Newtonian gravity theories whose algebras are obtained by expanding the 2D Euclidean versions of the so-called $\mathfrak{B}_{k}$ algebras \cite{Edelstein:2006se,Izaurieta:2009hz}. These are generalizations of the Maxwell algebra that have been useful to reproduce higher-dimensional standard General Relativity as a particular limit of generalized CS and Born-Infeld gravity theories \cite{Izaurieta:2009hz,Concha:2013uhq,Concha:2014zsa}. Interestingly, we show that the Carroll-Galilei gravity and its extension correspond to particular subcases of the post-Carroll-Newtonian gravity.

The paper is organized as follows: In section 2, we review the Chern-Simons formulation of a model based on the 2D Euclidean version of the AdS and Poincaré algebras, denoted in this work as AdS$_{2}^{\text{E}}$ and Poincaré$_{2}^{\text{E}}$, respectively. In section 3, we expand these algebras and we construct Carroll gravity models defined in three space-time dimensions based on them. In section 4, we show how to overcome the degeneracy problem that appears in the Carroll-Galilei gravity. An extended Carroll-Galilei gravity is found and then, a generalization to Post-Carroll-Newtonian extensions is considered by expanding a family of 2D Euclidean algebras.  Section 5 is devoted to conclusions and possible future developments.
\section{Euclidean AdS Chern-Simons theory}

In this section, we briefly review the construction of the CS theory based on the 2D Euclidean AdS algebra, which we have denoted as AdS$_{2}^{\mathrm{E}}$.  First, let us consider the commutation relations of the AdS$_{2}^{\mathrm{E}}$ algebra,
\begin{align}
\left[  \tilde{J},\tilde{P}_{a}\right]  &=\epsilon_{ab}\tilde{P}^{b}\,, &\left[  \tilde{P}_{a},\tilde{P}_{b}\right] & =-\epsilon_{ab}\tilde
{J}\,.\label{adse}%
\end{align}
Here the indices $a,b=1,2$ are raised and lowered with the Euclidean metric
$\delta_{ab}$, and the 2D Levi-Civita symbol is defined such that $\epsilon_{12}=\epsilon^{12}=1$. The generators $\tilde{P}_{a}$ generate spatial translations in
Euclidean bidimensional space, while $\tilde{J}$ generates one-dimensional
rotations. Note that AdS$_{2}^{\mathrm{E}}$ is isomorphic to the 3D Lorentz
algebra $\mathfrak{so}\left(  2,1\right)  $. Indeed, by defining $\tilde
{J}=J_{0}=-J^{0}$ and $J_a=\tilde{P}_{a}$ the commutation relations become
\begin{align}
\left[  J_{0},J_{a}\right]  &=\epsilon_{0ab}J
^{b}\,, &\left[  J_{a},J_{b}\right]
=\epsilon_{ab0}J^{0}\,.
\end{align}
Thus, the commutation relations (\ref{adse}) can be written in a Lorentz
covariant basis as $\left[  J_{A},J_{B}\right]  =\epsilon_{ABC}J^{C}$ for a
3D Lorentz index $A=\left(  0,a\right)  $ and the Minkowski
metric $\eta_{AB}=\mathrm{diag}\left(  -,+,+\right)  $. Furthermore, the
Lorentz rank-$2$ invariant tensor is given by $\left\langle J_{A}%
J_{B}\right\rangle =\tilde{\beta}\eta_{AB}$, where $\tilde{\beta}$ is an arbitrary constant.
In the AdS$_{2}^{\mathrm{E}}$ basis, the non-vanishing components of the invariant tensor
read
\begin{align}
\left\langle \tilde{P}_{a}\tilde{P}_{b}\right\rangle &=\tilde{\beta}\delta_{ab}\,, &
\left\langle \tilde
{J}\tilde
{J}\right\rangle &=-\tilde{\beta}\,.\label{inv1}%
\end{align}
Given a gauge connection one-form $A=A_{\mu}^{A}T_{A}\otimes\mathrm{d}x^{\mu}$
evaluated in a gauge Lie algebra, the
3D CS action principle is given by \footnote{Henceforth, the wedge product $\wedge$ between differential forms is implied.}
\begin{equation}
S_{\mathrm{CS}}\left[A\right]=\frac{k}{4\pi}\int_{\mathcal{M}}\left\langle A\mathrm{d}A+\frac{2}{3}%
A^{3}\right\rangle\,,\label{csgeneral}%
\end{equation}
where $\left\langle \dotsi\right\rangle $ denotes the symmetrized trace
acting on the generators $T_{A}$ and $k$ is the Chern-Simons level. With the purpose of writing down the CS action
for AdS$_{2}^{\mathrm{E}}$, we introduce the gauge connection evaluated in the
AdS$_{2}^{\mathrm{E}}$ algebra%
\begin{equation}
A=\tilde{\omega}\tilde{J}+\tilde{e}^{a}\tilde{P}_{a}\,.\label{connection1}%
\end{equation}
The corresponding gauge curvature two-form is given by $F=\mathrm{d}A+\frac
{1}{2}\left[  A,A\right]  $. We denote its components in the AdS$_{2}%
^{\mathrm{E}}$ basis as
\begin{equation}
F=R\left(  \tilde{\omega}\right)  \tilde{J}+R\left(  \tilde{e}^{a}\right)
\tilde{P}_{a}\,,
\end{equation}
where%
\begin{eqnarray}
R\left(  \tilde{\omega}\right)  &=&\mathrm{d}\tilde{\omega}-\frac{1}{2}%
\epsilon_{ab}\tilde{e}^{a}\tilde{e}^{b}\,,  \notag \\
R\left(  \tilde{e}%
^{a}\right)  &=&\mathrm{d}\tilde{e}^{a}-\epsilon^{ab}\tilde{\omega}\tilde{e}%
_{b}\,.\label{curvature1}%
\end{eqnarray}
From (\ref{csgeneral}) it is straightforward to show that the CS action for
AdS$_{2}^{\mathrm{E}}$ is given by
\begin{equation}
S_{\mathrm{CS}}^{\mathrm{AdS}_{2}^{\mathrm{E}}}=\frac{k\tilde{\beta}}{4\pi}\int\left(
R\left(  \tilde{e}^{a}\right)  \tilde{e}_{a}-\tilde{\omega}\mathrm{d}%
\tilde{\omega}\right)\,.\label{cs1}%
\end{equation}
The CS action \eqref{cs1} is equivalent to the 3D exotic Lagrangian for the $\mathrm{SO}\left(
2,1\right)  $ group when it is  written in terms of a Lorentz covariant basis \cite{Witten:1988hc,Zanelli:2005sa}.
Furthermore, the non-degeneracy of the $\mathfrak{so}\left(  2,1\right)  $
invariant tensor implies that the equations of motion emerging from the field variation of \eqref{adse} are given by the vanishing of the gauge curvatures in \eqref{curvature1}.

The writing of the $\mathfrak{so}\left(  2,1\right)  $ algebra in the
AdS$_{2}^{\mathrm{E}}$ basis implies a different interpretation in which the
generators $\tilde{P}_{a}$ are identified as translations. Consequently, the abelianization of this translation subspace leads to a Euclidean version of the Poincar\'{e} algebra, which we will refer to as Poincaré$_{2}^{\mathrm{E}}$, isomorphic to $\mathfrak{iso}\left( 2\right)  $.
This is performed by means of the following IW contraction: we rescale the
translation generators as $\tilde{P}_{a}\rightarrow\sigma\tilde{P}_{a}$, and then we apply the limit $\sigma\rightarrow\infty$, which can be seen as the vanishing cosmological constant one. The resulting commutation relations and non-vanishing components of the rank-$2$ invariant tensor are directly obtained by applying these rescaling and limit in (\ref{adse}) and (\ref{inv1}), leading to
\begin{align}
\left[  \tilde{J},\tilde{P}_{a}\right] & =\epsilon_{ab}\tilde{P}^{b}\,, &\left\langle \tilde{J}\tilde{J}\right\rangle &=-\tilde{\beta}\,.\label{inv3}%
\end{align}
The corresponding 3D CS action can be obtained by replacing the invariant tensor \eqref{inv3} and the Poincaré gauge connection one-form, which coincides with \eqref{connection1}, in the general CS expression \eqref{csgeneral}. Then, one finds that the CS action reads
\begin{equation}
S_{\mathrm{CS}}^{\mathrm{Poincar\acute{e}}_{2}^{\mathrm{E}}}=-\frac{k\tilde{\beta}
}{4\pi}\int\tilde{\omega}\mathrm{d}\tilde{\omega}\,.\label{cs3}%
\end{equation}
In this case, the components of the gauge curvature are given by
\begin{align}
R\left(  \tilde{\omega}\right)  &=\mathrm{d}\tilde{\omega}, &R\left(  \tilde{e}^{a}\right) & =\mathrm{d}\tilde{e}^{a}-\epsilon
^{ab}\tilde{\omega}\tilde{e}_{b}\,.
\end{align}
However, in contrast to the previous case, the lack of kinetic terms for the gauge fields $\tilde{e}^{a}$ associated with the translations in the action implies that it does not lead to the vanishing of the curvatures as equations of motion. In fact, from (\ref{cs3}) it is immediately seen that only $R\left(  \tilde{\omega
}\right)  $ is on-shell vanishing. This is a consequence of the degeneracy of the invariant tensor after applying the IW contraction.

Note that the CS action \eqref{cs3} can alternatively be recovered from the CS action for AdS$_{2}^{\mathrm{E}}$ \eqref{cs1} after a contraction process. Indeed, by rescaling the zweibein as follows:
\begin{align}
    \tilde{e}^{a}&\rightarrow\sigma^{-1} \tilde{e}^{a} \,,
\end{align}
the CS action \eqref{cs3} is recovered after performing the limit $\sigma\rightarrow\infty$.

\section{Carrollian gravity from 2D symmetry}

In this section, we will show that the AdS$_{2}^{\mathrm{E}}$ algebra as well as its Poincaré limit can be expanded to find Carrollian gravity models defined in three space-time dimensions. In particular, we focus on the derivation of AdS-Carroll gravity and the Carroll-Galilei one\footnote{Also referred as para-Galilei in \cite{Bacry:1968zf} or AdS-Static in \cite{Concha:2023bly,Concha:2024dap}.} from 2D symmetry. To this end, the $S$-expansion method \cite{Izaurieta:2006zz} will be applied with specific considerations on the semigroups.

\subsection{AdS-Carroll gravity}

Let us now consider the obtaining of the AdS-Carroll algebra, defined in three space-time dimensions,by means of the $S$-expansion method \cite{Izaurieta:2006zz}. Specifically, we consider a $S_{E}^{\left(  1\right)}$-expansion and $0_{S}$-reduction of AdS$_{2}^{\mathrm{E}}$. It is convenient
to recall that the semigroup $S_{E}^{\left(  1\right)  }=\left\{  \lambda
_{0},\lambda_{1},\lambda_{2}\right\}  $ presents the following multiplication
rule:%
\begin{equation}
\lambda_{\alpha}\lambda_{\beta}=\left\{
\begin{array}
[c]{ll}%
\lambda_{\alpha+\beta}\text{,} & \text{if }\alpha+\beta\leq1\text{,}\\
\lambda_{2}\text{,} & \text{otherwise.}%
\end{array}
\right.  \label{rule}%
\end{equation}
From (\ref{rule}), it follows that $\lambda_{2}$ is the zero element of the
semigroup. Then, the AdS-Carroll algebra appears after performing the $S_{E}^{\left(  1\right)  }$-expansion and
extracting the $0_{S}$-reduced algebra, namely $0_{S}\tilde{J}=0_{S}\tilde{P}_{a}=0$. One finds that the AdS-Carroll generators are related to the AdS$_{2}^{\mathrm{E}}$ ones in terms of the semigroup elements as
\begin{align}
J&=\lambda_{0}\tilde{J}\,, & H&=\lambda_{1}\tilde{J}\,, &  P_{a}&=\lambda_{0}\tilde{P}_{a}\,, & G_{a}&=\lambda
_{1}\tilde{P}_{a}\,,\label{exp1}%
\end{align}
and satisfy the following commutators%
\begin{align}
\left[  J,P_{a}\right]   &  =\epsilon_{ab}P^{b}, & \left[  J,G_{a}\right]   &
=\epsilon_{ab}G^{b}, & \left[  P_{a},P_{b}\right]   &  =-\epsilon
_{ab}J,\nonumber\\
\left[  P_{a},G_{b}\right]   &  =-\epsilon_{ab}H, & \left[  H,P_{a}\right]
&  =\epsilon_{ab}G^{b}\,. &  & \label{adsc}%
\end{align}
This is the 3D AdS-Carroll algebra. The corresponding invariant tensor can be
obtained directly by applying the S-expansion procedure on \eqref{inv1}:%
\begin{align}
\left\langle P_{a}P_{b}\right\rangle &=\beta_{0}\delta_{ab}\,, & \left\langle P_{a}G_{b}\right\rangle &=\beta_{1}\delta_{ab}\,, & \left\langle JJ\right\rangle &=-\beta_{0}\,, &\left\langle
JH\right\rangle &=-\beta_{1}\,,\label{inv2}%
\end{align}
where $\beta_{0}$ and $\beta_{1}$ are independent constants obtained from
\eqref{inv1} as $\beta_{0}=\lambda_{0}\tilde{\beta}$ and $\beta_{1}=\lambda_{1}\tilde{\beta}$.
The CS action for the AdS-Carroll algebra can be obtained by introducing the
one-form gauge connection evaluated in the gauge algebra:
\begin{equation}
A=\tau H+e^{a}P_{a}+\omega J+\omega^{a}G_{a}\,.\label{Aadsc}%
\end{equation}
The corresponding gauge curvature is given by
\begin{equation}
F=R\left(  \tau\right)  H+R\left(  e^{a}\right)  P_{a}+R\left(  \omega\right)
J+R\left(  \omega^{a}\right)  G_{a}\,,
\end{equation}
with%
\begin{align}
R\left(  \omega\right)   &  =\mathrm{d}\omega-\frac{1}{2}\epsilon_{ab}%
e^{a}e^{b}\,,\nonumber\\
R\left(  \tau\right)   &  =\mathrm{d}\tau-\epsilon_{ab}e^{a}%
\omega^{b}\,,\nonumber\\
R\left(  e^{a}\right)   &  =\mathrm{d}e^{a}-\epsilon^{ab}\omega e_{b}%
\,,\nonumber\\
R\left(  \omega^{a}\right)   &  =\mathrm{d}\omega^{a}-\epsilon^{ab}%
\omega\omega_{b}-\epsilon^{ab}\tau e_{b}\,.\label{curv2}%
\end{align}
The corresponding CS action for the AdS-Carroll algebra reads \cite{Matulich:2019cdo,Ravera:2019ize,Concha:2023bly,Aviles:2024llx}
\begin{equation}
S_{\mathrm{CS}}^{\mathrm{AdS}\text{\textrm{-}}\mathrm{Carroll}}=\frac{k}{4\pi
}\int\beta_{0}\left( R\left(  e^{a}\right)  e_{a}-\omega\mathrm{d}%
\omega\right)  +\beta_{1}\left(  R\left(  \omega^{a}\right)  e_{a}+R\left(
e^{a}\right)  \omega_{a}-2\omega\mathrm{d}\tau\right)\,. \label{csadsc}
\end{equation}
The CS action \eqref{csadsc} describes the most general action invariant under the AdS-Carroll algebra and contains two independent sectors. One the one hand, we have the exotic ultra-relativistic term along $\beta_0$. On the other hand, the term proportional to $\beta_1$ reproduces the standard AdS-Carroll gravity term which is isomorphic to the relativistic $\mathfrak{iso}\left(2,1\right)$ gravity Lagrangian by interchanging the role of $\omega^{a}$ and $e^{a}$. As it happened in the previous case, the non-degeneracy of the invariant tensor
\eqref{inv2} implies that the equations of motion are given by vanishing of
the gauge curvatures \eqref{curv2}. Let us note that a vanishing cosmological constant limit $\ell\rightarrow\infty$ can be applied to the AdS-Carroll algebra \eqref{adsc} after considering the following space-time rescaling of the generators:
\begin{align}
    H&\rightarrow \ell H\,, & P_{a}&\rightarrow \ell P_{a}\,.\label{resc}
\end{align}
Then, the flat limit reproduces the Carroll algebra and the CS gravity action is now invariant by construction under the Carroll group. In particular, the flat limit $\ell\rightarrow\infty$ can be applied directly to the CS action \eqref{csadsc} after considering the rescaling of the gauge fields:
\begin{align}
\tau&\rightarrow \ell^{-1}\tau\,, & e^{a}&\rightarrow \ell^{-1}e^{a}\,,   
\end{align}
along with
\begin{equation}
    \beta_{1}\rightarrow\ell\beta_1\,.
\end{equation}
\subsection{Carroll-Galilei gravity}

Let us now consider the $S_{E}^{\left(  1\right)  }$-expansion and $0_{S}
$-reduction of $\mathfrak{iso}\left(  2\right)$ following the same procedure considered to obtain the AdS-Carroll algebra from AdS$_{2}^{\mathrm{E}}$. By expanding the generators of $\mathfrak{iso}\left(
2\right)  $ according to (\ref{exp1}), one finds that the $0_{S}$-reduced
$S_{E}^{\left(  1\right)}$-expanded algebra corresponds to the 3D Carroll-Galilei algebra which we have denoted as $\mathfrak{car}$-$\mathfrak{gal}$. The $\mathfrak{car}$-$\mathfrak{gal}$ commutation relations are given by
\begin{align}
\left[  J,P_{a}\right]  &=\epsilon_{ab}P^{b}\,, &\left[
J,G_{a}\right]  &=\epsilon_{ab}G^{b}\,, & \left[  H,P_{a}\right]
&=\epsilon_{ab}G^{b}\,.\label{CG}%
\end{align}
Note that it can also be obtained from the AdS-Carroll algebra \eqref{adsc} through an IW contraction. In fact, by rescaling the AdS-Carroll generators as $P_{a}\rightarrow\varepsilon P_{a}%
$, $G_{b}\rightarrow\varepsilon G_{a}$ and taking the limit $\varepsilon
\rightarrow\infty$, one recovers the Carroll-Galilei commutation relations \eqref{CG}. Such contraction process corresponds to the non-relativistic limit in which the speed of light $c$ tends to infinity \footnote{Here the $\varepsilon$ parameter can be identified with the speed of light $\varepsilon\sim c$.}. The so-called static algebra \cite{Bacry:1968zf} can be obtained from the Carroll-Galilei algebra \eqref{CG} after considering the space-time rescaling of the generators \eqref{resc} and performing the limit of the vanishing cosmological constant $\ell\rightarrow\infty$. The static algebra can alternatively be obtained from the Carroll one as a non-relativistic contraction.

The rank-$2$ invariant tensor can be obtained from \eqref{inv3} by means of the
same expansion procedure, resulting in the following components:%
\begin{align}
\left\langle JJ\right\rangle &=-\mu_{0}\,, & \left\langle
JH\right\rangle &=-\mu_{1}\,,\label{inv4}%
\end{align}
where the new constants are obtained from the original one through
$\mu_{0}=\lambda_{0}\tilde{\beta}$ and $\mu_{1}=\lambda_{1}\tilde{\beta}$. 

Let us now introduce the one-form gauge connection evaluated in the Carroll-Galilei algebra:
\begin{equation}
A=\tau H+e^{a}P_{a}+\omega J+\omega^{a}G_{a}\,.\label{1fcg}
\end{equation}
The corresponding components of the curvature two-form read
\begin{align}
R\left(  \omega\right)   &  =\mathrm{d}\omega\,,\nonumber\\
R\left(  \tau\right)   &  =\mathrm{d}\tau\,,\nonumber\\
R\left(  e^{a}\right)   &  =\mathrm{d}e^{a}-\epsilon^{ab}\omega e_{b}\,,\nonumber\\
R\left(  \omega^{a}\right)   &  =\mathrm{d}\omega^{a}-\epsilon^{ab}%
\omega\omega_{b}-\epsilon^{ab}\tau e_{b}\,.
\end{align}
Then, by replacing the gauge connection one-form \eqref{1fcg} and the invariant tensor \eqref{inv4} in the CS general expression \eqref{csgeneral}, we obtain the following CS action:
\begin{equation}
S_{\mathrm{CS}}^{\mathfrak{car}\text{-}\mathfrak{gal}}=-\frac{k}{4\pi}\int
\mu_{0}\omega R(\omega)+2\mu_{1}\tau R(\omega ) \,. \label{cscargal}
\end{equation}
Similar to the $\mathfrak{iso}\left(2\right)$ CS action \eqref{cs3}, this action does not include kinetic terms for $e^{a}$
and $\omega^{a}$. In both cases, the degeneracy of the invariant tensor
prevents the on-shell vanishing of the curvature two-forms.

\begin{center}
 \begin{figure}[h!]
  \begin{center}
        \includegraphics[width=9.6cm, height=5.5cm]{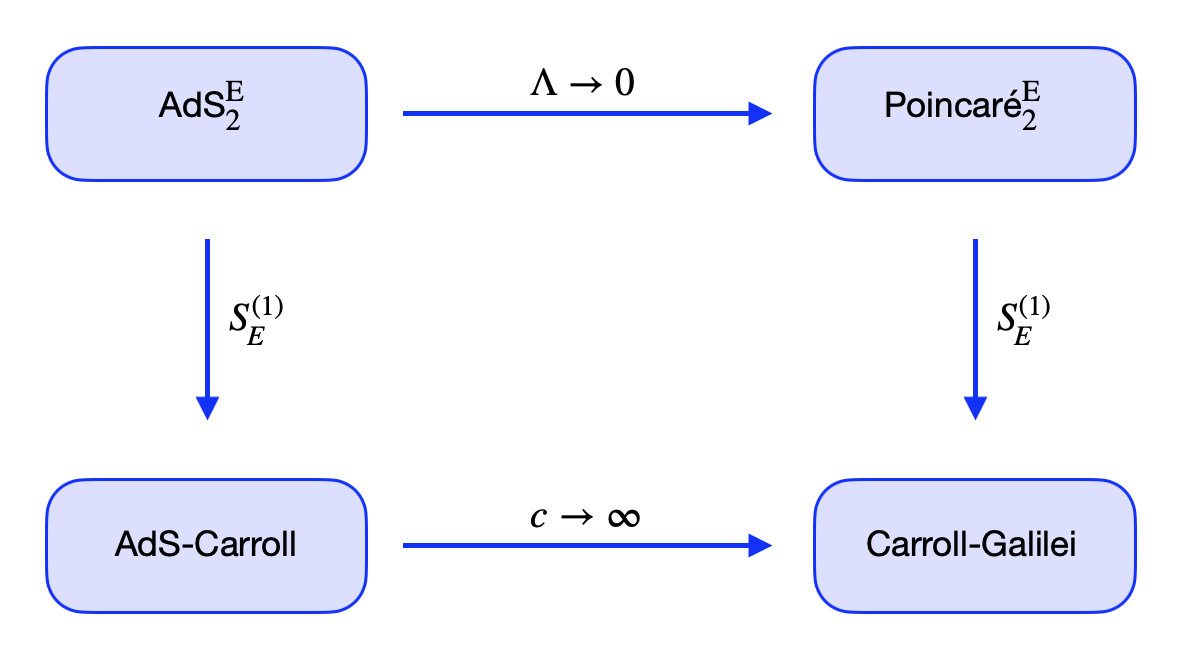}
        \captionsetup{font=footnotesize}
        \caption{Diagram summarizing the S-expansion, flat limit and non-relativistic limit.} 
        \label{fig1}
         \end{center}
        \end{figure}
    \end{center}

As an ending remark, let us note that the vanishing cosmological constant limit $\Lambda\rightarrow 0$ applied to the 2D Euclidean AdS algebra to obtain the $\mathfrak{iso}\left(2\right)$ algebra can be seen as a non-relativistic limit $c\rightarrow\infty$ in the expanded sector defined on three space-time dimensions (see figure \ref{fig1}). This is similar to what happens in the surprising duality between the 2D conformal Galilei algebra \cite{Bagchi:2009my,Bagchi:2009pe} and the asymptotic symmetry of 3D flat gravity \cite{Barnich:2006av,Barnich:2010eb,Barnich:2012aw} ( $\mathfrak{cga}/\mathfrak{bms}_{3}$ duality \cite{Bagchi:2010zz}) in which the flat limit applied to the conformal algebra to derive the $\mathfrak{bms}_{3}$ algebra can be interpreted as a non-relativistic limit. Note that the non-relativisitc limit applied to the AdS-Carroll symmetry reproducing the Carroll-Galilei algebra was already evidenced in the original cube of Bacry and Lévy-Leblond \cite{Bacry:1968zf}. This limit can be seen as the non-relativistic limit of the Poincaré algebra leading to the usual Galilei algebra, since the AdS-Carroll and the Carroll-Galilei algebras are isomorphic to the Poincaré and Galilei ones, respectively \cite{Bacry:1968zf}.

\section{Extended Carroll-Galilei gravity from 2D symmetry}

In this section, we show that a central extension of the Carroll-Galilei algebra with a non-degenerate bilinear trace can be obtained from a particular 2D symmetry. Indeed, the degeneracy issue appearing in the previous section in both the Euclidean sector and the expanded one can be solved by adding a central charge to the Euclidean 2D Poincaré algebra:
\begin{align}
\left[  \tilde{J},\tilde{P}_{a}\right]&=\epsilon_{ab}\tilde{P}^{b}\,, &  \left[  \tilde{P}_{a},\tilde{P}_{b}\right]&=-\epsilon_{ab}\tilde
{Z}\,.\label{eiso}%
\end{align}
The 2D algebra \eqref{eiso} can be seen as the Euclidean 2D version of the Maxwell algebra \cite{Bacry:1970ye,Bacry:1970du,Schrader:1972zd,Gomis:2017cmt}, which we shall denote as Maxwell$_{2}^{\mathrm{E}}$. As it was discussed in \cite{Matulich:2019cdo} and previously noted in \cite{Soroka:2004fj}, the Maxwell algebra corresponds to the most natural extension of Poincaré with a non-degenerate invariant metric. In particular, the Maxwell$_{2}^{\mathrm{E}}$ admits the following non-vanishing components
of the rank-$2$ invariant tensor \cite{Salgado:2014jka}:
\begin{align}
\left\langle \tilde{P}_{a}\tilde{P}_{b}\right\rangle &=\tilde{\beta}_{2}\delta_{ab}\,, & \left\langle \tilde{J}\tilde{J}\right\rangle &=-\tilde{\beta}_{0}\,,  & \left\langle \tilde{J}\tilde{Z}\right\rangle &=-\tilde{\beta}_{2}\,,\label{inv1b}%
\end{align}
with $\tilde{\beta}_{0}$ and $\tilde{\beta}_{2}$ being arbitrary constants.

Let us now consider the gauge connection one-form for Maxwell$_{2}^{\mathrm{E}}$
\begin{equation}
A=\tilde{\omega}\tilde{J}+\tilde{e}^{a}\tilde{P}_{a}+\tilde{k}\tilde{Z}\,,\label{Amax}%
\end{equation}
along its corresponding gauge curvature two-form 
\begin{equation}
F=R\left(  \tilde{\omega}\right)  \tilde{J}+R\left(  \tilde{k}\right)  \tilde
{Z}+R\left(  \tilde{e}^{a}\right)  \tilde{P}_{a}\,,
\end{equation}
with
\begin{align}
R\left(  \tilde{\omega}\right)   &  =\mathrm{d}\tilde{\omega}\,,\nonumber\\
R\left(  \tilde{k}\right)   &  =\mathrm{d}\tilde{k}-\frac{1}{2}\epsilon_{ab}
\tilde{e}^{a}\tilde{e}^{b}\,,\nonumber\\
R\left(  \tilde{e}^{a}\right)   &  =\mathrm{d}\tilde{e}^{a}-\epsilon^{ab}
\tilde{\omega}\tilde{e}_{b}\,.\label{2fMax}
\end{align}
By replacing the Maxwell$_{2}^{\mathrm{E}}$ gauge connection
\eqref{Amax}, the commutation relations \eqref{eiso} and the invariant tensors
\eqref{inv1b} into the general expression \eqref{csgeneral}, we find that the
3D CS action for the Maxwell$_{2}^{\mathrm{E}}$ algebra is given by
\begin{equation}
S_{\mathrm{CS}}^{\text{\textrm{Maxwell}}_{2}^{\mathrm{E}}}=\frac{k}{4\pi}%
\int \tilde{\beta}_{2}\left(  R\left(  \tilde{e}^{a}\right)  \tilde{e}_{a}%
-2\tilde{k}R\left(  \tilde{\omega}\right)\right)
-\tilde{\beta}_{0}\tilde{\omega}R\left(  \tilde{\omega}\right)\,.\label{CSMax2}
\end{equation}
Unlike the action of Poincar\'{e}$_{2}^{\mathrm{E}}$, the non-degeneracy of the invariant bilinear trace \eqref{inv1b} ensures that previous action presents a kinetic term for each gauge field. In particular, the non-degeneracy condition requires $\tilde{\beta}_{2}\neq 0$ which implies that the field equations are given by the vanishing of the curvature two-forms \eqref{2fMax}. Let us note that the CS term appearing along $\tilde{\beta}_{2}$ \eqref{CSMax2} coincides with the Nappi-Witten term defined in \cite{Concha:2020eam}. This is not a surprise since the Maxwell$_{2}^{\mathrm{E}}$ algebra \cite{Afshar:2019axx,Salgado-Rebolledo:2021wtf} is isomorphic to the Nappi-Witten algebra \cite{Nappi:1993ie,Figueroa-OFarrill:1999cmq}.

\subsection{Extended Carroll-Galilei gravity}

Let us now consider a $0_{S}$-reduced $S_{E}^{\left(  1\right)  }$-expansion of the
Maxwell$_{2}^{\mathrm{E}}$ algebra \eqref{eiso}. The resulting algebra appears after extracting the $0_{S}$-reduced algebra from the $S_{E}^{\left(  1\right)  }$-expanded one. The new generators are related to the Maxwell$_{2}^{\mathrm{E}}$ ones through the semigroup elements as follows:
\begin{align}
    J&=\lambda_0 \tilde{J}\,, &  P_{a}&=\lambda_0 \tilde{P}_{a}\,, & S&=\lambda_0 \tilde{Z}\,,\notag\\
    H&=\lambda_1 \tilde{J}\,, &  G_{a}&=\lambda_1 \tilde{P}_{a}\,, & M&=\lambda_1 \tilde{Z}\,, 
\end{align}
and satisfy the following commutation relations
\begin{align}
\left[  J,P_{a}\right]   &  =\epsilon_{ab}P^{b}, & \left[  J,G_{a}\right]   &
=\epsilon_{ab}G^{b}, & \left[  H,P_{a}\right]   &  =\epsilon_{ab}%
G^{b}\,,\nonumber\\
\left[  G_{a},P_{b}\right]   &  =-\epsilon_{ab}M, & \left[  P_{a}%
,P_{b}\right]   &  =-\epsilon_{ab}S\,. &  &
\end{align}
We refer to this algebra as the extended Carroll-Galilei algebra\footnote{Also denoted as extended para-Bargmann in \cite{Concha:2023bly} and extended AdS-static in \cite{Concha:2024dap}.}, which we have denoted as the extended $\mathfrak{car}$-$\mathfrak{gal}$ algebra. The
expansion procedure also provides us with the non-vanishing components of the rank-$2$ invariant tensor:
\begin{align}
\left\langle P_{a}G_{b}\right\rangle  &  =\gamma_{1}\delta_{ab}\,, &
\left\langle P_{a}P_{b}\right\rangle  &  =\gamma_{0}\delta_{ab}\,, &
\left\langle JJ\right\rangle  &  =-\mu_{0}\,, & \left\langle JH\right\rangle
&  =-\mu_{1}\,,\nonumber\\
\left\langle JS\right\rangle  &  =-\gamma_{0}\,, & \left\langle JM\right\rangle
&  =-\gamma_{1}\,, & \left\langle HS\right\rangle  &  =-\gamma_{1}\,, &  &
\end{align}
where $\gamma_{\alpha}=\lambda_{\alpha}\tilde{\beta}_{2}$
and $\mu_{\alpha}=\lambda_{\alpha}\tilde{\beta}_{0}$.

The CS action can be constructed from the gauge connection one-form for the extended Carroll-Galilei algebra:
\begin{equation}
A=\omega J+\tau H+sS+mM+\omega ^{a}G_{a}+e^{a}P_{a}\,.
\end{equation}%
The corresponding gauge curvature two-form is given by%
\begin{equation}
F=R\left( \omega \right) J+R\left( \tau \right) H+R\left( s\right) S+R\left(
m\right) M+R\left( \omega ^{a}\right) G_{a}+R\left( e^{a}\right) P_{a}\,,
\end{equation}%
with%
\begin{align}
R\left( \omega \right) & =\mathrm{d}\omega\,,  \notag \\
R\left( \tau \right) & =\mathrm{d}\tau\,,  \notag \\
R\left( s\right) & =\mathrm{d}s-\frac{1}{2}\epsilon _{ab}e^{a}e^{b}\,,
\notag \\
R\left( m\right) & =\mathrm{d}m-\epsilon _{ab}\omega ^{a}e^{b}\,,  \notag \\
R\left( \omega ^{a}\right) & =\mathrm{d}\omega ^{a}-\epsilon ^{ab}\omega
\omega _{b}-\epsilon ^{ab}\tau e_{b}\,,  \notag \\
R\left( e^{a}\right) & =\mathrm{d}e^{a}-\epsilon ^{ab}\omega e_{b}\,.\label{2fecargal}
\end{align}%
Thus, the most general CS action for the extended Carroll-Galilei algebra reads
\begin{align}
S_{\mathrm{CS}}^{\text{\textrm{Extended}}\, \mathfrak{car}\text{-}\mathfrak{gal}}& = S_{\mathrm{CS}}^{\gamma\text{-}\, \mathfrak{car}\text{-}\mathfrak{gal}}+S_{\mathrm{CS}}^{\mathfrak{car}\text{-}\mathfrak{gal}}\,, \label{csecg}
\end{align}
where $S_{\mathrm{CS}}^{\mathfrak{car}\text{-}\mathfrak{gal}}$ is the Carroll-Galilei CS action \eqref{cscargal} and
\begin{align}
S_{\mathrm{CS}}^{\gamma\text{-}\, \mathfrak{car}\text{-}\mathfrak{gal}}& =\frac{k}{4\pi }\int \gamma_{0}\left( R\left( e^{a}\right) e_{a}-2sR\left(  \omega\right)\right)   \notag \\
& +\gamma_{1}\left( R\left( e^{a}\right) \omega _{a}+R\left( \omega^{a}\right) e_{a}-2sR\left( \tau \right)-2mR\left( \omega \right) \right)\,. \label{csecg1}
\end{align}
Let us note that the CS action \eqref{csecg} containing the Carrol-Galilei CS term is invariant under the extended Carroll-Galilei algebra. However one can omit the Carroll-Galilei CS term since the non-degeneracy criteria requires $\gamma_1\neq 0$. The CS action proportional to the $\gamma$'s contains two sectors: an exotic one along $\gamma_0$ and the $\gamma_1$ term. The latter is isomorphic to the extended Bargmann CS action \cite{Papageorgiou:2009zc,Bergshoeff:2016lwr} by interchanging the role of the $\omega^{a}$ and $e^{a}$ gauge fields. Nonetheless, the field equations differ from the extended Bargmann ones since they are given by the vanishing of the extended Carroll-Galilei curvature two-forms \eqref{2fecargal}. Note that the CS action for the extended static algebra \cite{Concha:2023bly} can be recovered from the extended Carroll-Galilei one \eqref{csecg1} after considering the following rescaling ,
\begin{align}
\tau&\rightarrow \ell^{-1}\tau\,, & e^{a}&\rightarrow \ell^{-1}e^{a}\,, &    \gamma_{1}&\rightarrow\ell\gamma_1\,,
\end{align}
and performing the vanishing cosmological constant limit $\ell\rightarrow\infty$.
\subsection{Post-Carroll-Newtonian generalization}
The generalization of our results to Post-Carroll-Newtonian extensions can be performed by expanding a family of 2D Euclidean algebras. Indeed, one can show that the Poincaré$_{2}^{\mathrm{E}}$ and Maxwell$_{2}^{\mathrm{E}}$ algebras belong to a larger family of Euclidean symmetries: the Euclidean 2D version of the $\mathfrak{B}_{k}$ algebras \cite{Edelstein:2006se,Izaurieta:2009hz}. This algebra is spanned by $\{\tilde{J}^{\left(i\right)},\tilde{P}_{a}^{\left(p\right)}\}$ with 
\begin{eqnarray}
    i&=&0,1,\cdots,\left[\frac{k-2}{2}\right]\,,\notag\\
    p&=&1,2,\cdots,\left[\frac{k-1}{2}\right]\,.
\end{eqnarray} 
Here, $\left[\,\cdot\,\right]$ denotes the integer part. These generators satisfy the following commutation relations:
\begin{align}
    \left[  \tilde{J}^{\left(i\right)},\tilde{P}_{a}^{\left(p\right)}\right]&=\epsilon_{ab}\tilde{P}^{b\,\left(i+p\right)}\,, &  \left[  \tilde{P}_{a}^{\left(p\right)},\tilde{P}_{b}^{\left(q\right)}\right]&=-\epsilon_{ab}\tilde
{J}^{\left(p+q-1\right)}\,. \label{EBK}
\end{align}
Let us note that for $k=3$ and $k=4$, we recover the Poincaré$_{2}^{\mathrm{E}}$ and the Maxwell$_{2}^{\mathrm{E}}$ algebras, respectively.  The 2D $\mathfrak{B}_{k}$ algebra admits a non-degenerate bilinear invariant trace only for even values of $k$:
\begin{align}
    \left\langle \tilde{P}_{a}^{\left(p\right)}\tilde{P}_{b}^{\left(q\right)}\right\rangle &=\tilde{\beta}_{p+q}\delta_{ab}\,, & \left\langle \tilde{J}^{\left(i\right)}\tilde{J}^{\left(j\right)}\right\rangle &=-\tilde{\beta}_{2\left(i+j\right)}\,.
\end{align}
Indeed, the non-degeneracy criteria requires that both components of the invariant tensor coexist which occurs for $k=4$ and in general for even values of $k$.

Post-Carroll-Newtonian algebras can be obtained after considering a $0_S$-reduced $S_{E}^{\left(1\right)}$-expansion of the 2D $\mathfrak{B}_{k}$ algebra \eqref{EBK}. The expanded generators are related to the 2D $\mathfrak{B}_{k}$ ones through the semigroup elements as follows:
\begin{align}
    J^{\left(i\right)}&=\lambda_0 \tilde{J}^{\left(i\right)}\,, &  P_{a}^{\left(p\right)}&=\lambda_0 \tilde{P}_{a}^{\left(p\right)}\,, \notag\\
    H^{\left(i\right)}&=\lambda_1 \tilde{J}^{\left(i\right)}\,, &  G_{a}^{\left(p\right)}&=\lambda_1 \tilde{P}_{a}^{\left(p\right)}\,. 
\end{align}
Then, by considering the original commutators of the 2D $\mathfrak{B}_{k}$ algebra \eqref{EBK} along the multiplication law of the $S_{E}^{\left(1\right)}$ semigroup \eqref{rule},  we find that the expanded generators satisfy
\begin{align}
    \left[  J^{\left(i\right)},P_{a}^{\left(p\right)}\right]   &  =\epsilon_{ab}P^{b\,\left(i+p\right)}, & \left[  J^{\left(i\right)},G_{a}^{\left(p\right)}\right]   &
=\epsilon_{ab}G^{b\,\left(i+p\right)}, & \left[  H^{\left(i\right)},P_{a}^{\left(p\right)}\right]   &  =\epsilon_{ab}%
G^{b\,\left(i+p\right)}\,,\nonumber\\
\left[  G_{a}^{\left(p\right)},P_{b}^{\left(q\right)}\right]   &  =-\epsilon_{ab}H^{\left(p+q-1\right)}, & \left[  P_{a}^{\left(p\right)}%
,P_{b}^{\left(q\right)}\right]   &  =-\epsilon_{ab}J^{\left(p+q-1\right)}\,. &  & \label{pcn}
\end{align}
The new algebra \eqref{pcn} corresponds to a post-Carroll-Newtonian algebra which we have denoted as $\mathfrak{pcn}_{m}$ and reproduces the Carroll-Galilei algebra and the extended Carroll-Galilei one for $m=1$ and $m=2$, respectively. In general, the expansion of the 2D $\mathfrak{B}_{k}$ algebra leads to $\mathfrak{pcn}_{m}$ with $m=k-2$. Interestingly, these algebras are isomorphic to the Post-Newtonian algebras \cite{Gomis:2019sqv,Gomis:2019nih,Concha:2023bly}  by interchanging $P_{a}^{\left(p\right)}$ and $G_{a}^{\left(p\right)}$. A flat limit $\ell\rightarrow\infty$ can be performed after considering the following rescaling of the generators:
\begin{align}
    P_{a}^{\left(p\right)}&\rightarrow \ell P_{a}^{\left(p\right)}\,, & H^{\left(i\right)}&\rightarrow\ell H^{\left(i\right)}.
\end{align}
The contracted algebra corresponds to a generalization of the static algebra\cite{Bacry:1968zf}. As in the 2D $\mathfrak{B}_{k}$ algebra, the $\mathfrak{pcn}_{m}$ algebra admits a non-degenerate invariant tensor only for even value of $m$:
\begin{align}
    \left\langle P_{a}^{\left(p\right)}G_{b}^{\left(q\right)}\right\rangle  &  =\beta_{p+q}\delta_{ab}\,, &
\left\langle P_{a}^{\left(p\right)}P_{b}^{\left(q\right)}\right\rangle  &  =\alpha_{p+q}\delta_{ab}\,, \notag \\
\left\langle J^{\left(i\right)}J^{\left(j\right)}\right\rangle  &  =-\alpha_{2\left(i+j\right)}\,, & \left\langle J^{\left(i\right)}H^{\left(j\right)}\right\rangle
&  =-\beta_{2\left(i+j\right)}\,,\label{invpcn}
\end{align}
where $\alpha_{n}=\lambda_0\tilde{\beta}_{n}$ and $\beta_{n}=\lambda_1\tilde{\beta}_{n}$. In particular, $\alpha_{n}$ is related to exotic CS terms.

The gauge connection one-form for the $\mathfrak{pcn}_{m}$ algebra reads
\begin{align}
    A&=\sum_{i=0}^{\left[\frac{m}{2}\right]}\left(\tau^{\left(i\right)} H^{\left(i\right)}+\omega^{\left(i\right)} J^{\left(i\right)}\right)+\sum_{p=1}^{\left[\frac{m+1}{2}\right]}\left(e^{a\,\left(p\right)}P_{a}^{\left(p\right)}+\omega^{a\,\left(p\right)}G_{a}^{\left(p\right)}\right)\,.\label{1fpcn}
\end{align}
Then, the CS action based on the $\mathfrak{pcn}_{m}$ algebra is obtained by considering the gauge connection one-form \eqref{1fpcn} and the invariant tensor \eqref{invpcn} into the general CS expression \eqref{csgeneral}:
\begin{align}
    S_{\mathrm{CS}}^{\mathfrak{pcn}_{m}}&=\frac{k}{4\pi}\int \sum_{n=1}^{2m-2}  \beta_n\,\left[2\delta_{p+q}^{n}\,e_a^{\left(p\right)}\mathcal{R}^{a}\left(\omega^{b \,\left(q\right)}\right)-2\delta_{2\left(i+j\right)}^{n}\,\tau^{\left(i\right)} \mathcal{R}\left(\omega^{\left(j\right)}\right)+\delta_{i+p+q}^{n}\epsilon^{ac}\tau^{\left(i\right)} e_{a}^{\left(p\right)}e_{c}^{\left(q\right)}\right]\,,\label{cspcgn}
\end{align}
where
\begin{align}
    \mathcal{R}\left(\omega^{\left(p\right)}\right)&=\mathrm{d}\omega^{\left(p\right)}\,, \notag \\
    \mathcal{R}^{a}\left(\omega^{b \,\left(p\right)}\right)&=\mathrm{d}\omega^{a \,\left(p\right)}-\sum_{i=0}^{\left[\frac{m}{2}\right]}\sum_{q=1}^{\left[\frac{m+1}{2}\right]}\delta_{i+q}^{p}\, \epsilon^{ac}\omega^{\left(i\right)}\omega_c^{\left(q\right)}\,.
\end{align}
Here, we have intentionally omitted the exotic terms along the $\alpha$'s since the non-degeneracy is not affected by setting $\alpha_{n}=0$. One can note that the Carroll-Galilei CS action \eqref{cscargal} is recovered for $m=1$ by considering $\beta_0=\mu_1$ together with the following identification of the gauge fields:
\begin{align}
    \tau^{\left(0\right)}&=\tau\,, & \omega^{\left(0\right)}&=\omega\,, & e^{a\,\left(1\right)}&=e^{a}\,, & \omega^{a\,\left(1\right)}&=\omega^{a}\,.
\end{align}
On the other hand, $m=2$ reproduces the extended Carroll-Galilei CS action \eqref{csecg} with $\beta_2=\gamma_1$, $\tau^{\left(1\right)}=m$ and $\omega^{\left(1\right)}=s$. Remarkably, the most general $\mathfrak{pcn}_{m}$ CS action can be written as a sum of CS actions,
\begin{eqnarray}
    S_{\mathrm{CS}}^{\mathfrak{pcn}_{m}}& = &S_{\mathrm{CS}}^{\mathfrak{car}\text{-}\mathfrak{gal}}+S_{\mathrm{CS}}^{\gamma\text{-}\, \mathfrak{car}\text{-}\mathfrak{gal}}+\sum_{n=3}^{m}S_{\mathrm{CS}}^{\mathfrak{pcn}_{n}}\,,
\end{eqnarray}
where each $S_{\mathrm{CS}}^{\mathfrak{pcn}_{n}}$ CS term is invariant under the $\mathfrak{pcn}_{n}$ algebra. Here, $S_{\mathrm{CS}}^{\mathfrak{car}\text{-}\mathfrak{gal}}$ denotes the Carroll-Galilei CS action \eqref{cscargal} and $S_{\mathrm{CS}}^{\gamma\text{-}\, \mathfrak{car}\text{-}\mathfrak{gal}}$ is the extended Carroll-Galilei CS term \eqref{csecg1}.
\section{Conclusion}
In this work we have shown that various Carroll gravity models in three space-time dimensions can be systematically derived from 2D Euclidean algebras using the $S$-expansion method \cite{Izaurieta:2006zz}. Starting with the 2D Euclidean AdS and Poincaré algebras, we obtained the AdS-Carroll and Carroll-Galilei algebras, and constructed the corresponding Chern-Simons gravity actions. Interestingly, analogous to the $\mathfrak{cga}/\mathfrak{bms}_{3}$ duality \cite{Bagchi:2010zz}, the vanishing cosmological constant limit applied to the 2D Euclidean AdS algebra translates into a non-relativistic limit in three space-time dimensions after the expansion.  However, both limits exhibit degeneracy that prevents the on-shell vanishing of the curvature two-forms as field equations. To address this issue, we introduced a central extension of the 2D Euclidean Poincaré algebra: the 2D Euclidean version of the Maxwell algebra \cite{Bacry:1970ye,Bacry:1970du,Schrader:1972zd,Gomis:2017cmt}. Expanding this algebra allowed us to obtain an extended Carroll-Galilei algebra equipped with a non-degenerate invariant tensor, enabling the construction of a well-defined CS action. Furthermore, we generalized our approach to Post-Carroll-Newtonian algebras, which can be derived by expanding the 2D Euclidean version of the $\mathfrak{B}_{k}$ algebra.

There are several generalizations of our results that would be worth exploring. Our approach could be extended to the supersymmetric case to construct Carroll CS supergravity models, along with their possible extensions. One might expect not only to recover the AdS-Carroll CS supergravity introduced in \cite{Ravera:2019ize,Ali:2019jjp,Concha:2024dap} but also to derive novel post-Carroll-Newtonian extensions. However, as noted in \cite{Concha:2021jnn}, the non-degeneracy issue becomes more subtle in the Maxwellian case. Similar to the non-relativistic limit, the degeneracy could potentially be resolved by including additional bosonic content.

Another natural extension of our results is to consider the coupling of higher-spin gauge fields. While Carroll CS gravity coupled to spin-3 gauge fields has been studied in \cite{Bergshoeff:2016soe,Concha:2022muu,Caroca:2022byi}, the construction of Carroll hypergravity with spin-$\frac{5}{2}$ gauge fields remains unexplored. The approach presented here could first be applied to derive the spin-$3$ Carroll CS gravity and subsequently be extended to construct the first example of Carroll hypergravity. Such a theory would correspond to the ultra-relativistic regime of known hypergravity theories \cite{Aragone:1983sz,Zinoviev:2014sza,Henneaux:2015tar,Henneaux:2015ywa,Fuentealba:2015jma,Fuentealba:2015wza}.

Finally, the isomorphism between various Carrollian and relativistic algebras offers a promising avenue for analyzing asymptotic symmetry. As pointed out in \cite{Aviles:2024llx}, the isomorphism between the AdS-Carroll algebra and the Poincaré one has been used to map 3D flat cosmology onto AdS-Carroll geometry. Moreover, they showed that the infinite-dimensional asymptotic symmetry algebra of AdS-Carroll CS gravity is given by the $\mathfrak{bms}_{3}$ algebra. It would be interesting to analyze the 3D Maxwell CS gravity case whose asymptotic symmetry algebra is given by a deformation and extension of the $\mathfrak{bms}_{3}$ algebra \cite{Concha:2018zeb}. The isomorphism between the Maxwell algebra and a Carrollian one \cite{Concha:2023bly} could provide valuable insights into the asymptotic symmetries of generalized Carrollian gravity models.

\section*{Acknowledgment}
P.C. acknowledges financial support from the National Agency for Research and
Development (ANID) through Fondecyt grants No. 1211077 and 11220328. E.R.
acknowledges financial support from ANID through SIA grant No. SA77210097
and Fondecyt grant No. 11220486. P.C. and E.R. would like to thank to the
Direcci\'{o}n de Investigaci\'{o}n and Vicerector\'{\i}a de Investigaci\'{o}%
n of the Universidad Cat\'{o}lica de la Sant\'{\i}sima Concepci\'{o}n,
Chile, for their constant support. S.S. acknowledges financial support from
Universidad de Tarapac\'{a}, Chile.

\bibliographystyle{fullsort}
 
\bibliography{Carroll_gravity_from_2D_Euclidean_symmetry}

\end{document}